\documentclass[11pt]{article}

\usepackage[margin=2.5cm]{geometry}
\usepackage[utf8]{inputenc}
\usepackage{amsmath}
\usepackage{graphicx}
\usepackage[sort&compress]{natbib}
\usepackage{hyperref}
\usepackage{url}

\title{pyABC: Efficient and robust easy-to-use approximate Bayesian computation}

\author{
Yannik Schälte\,$^{\text{1,2,3}}$, Emmanuel Klinger\,$^{\text{2,3,4}}$, Emad Alamoudi\,$^{\text{1}}$, and
Jan Hasenauer\,$^{\text{1,2,3},\ast}$
}

\date{}

\begin{document}

\maketitle
{\small
$^{\text{1}}$ Faculty of Mathematics and Natural Sciences, Rheinische Friedrich-Wilhelms-Universität Bonn, Bonn, Germany\\
$^{\text{2}}$ Institute of Computational Biology,
Helmholtz Zentrum München, Neuherberg, Germany\\
$^{\text{3}}$ Center for Mathematics, Technische Universität München, Garching, Germany\\
$^{\text{4}}$ Department of Connectomics, Max Planck Institute for Brain Research, Frankfurt, Germany\\
$^\ast$ To whom correspondence should be addressed (jan.hasenauer@uni-bonn.de)
}

\begin{abstract}
The Python package pyABC provides a framework for approximate Bayesian computation (ABC), a likelihood-free parameter inference method popular in many research areas.
At its core, it implements a sequential Monte-Carlo (SMC) scheme, with various algorithms to adapt to the problem structure and automatically tune hyperparameters.
To scale to computationally expensive problems, it provides efficient parallelization strategies for multi-core and distributed systems.
The package is highly modular and designed to be easily usable.
In this major update to pyABC, we implement several advanced algorithms that facilitate efficient and robust inference on a wide range of data and model types.
In particular, we implement algorithms to account for noise, to adaptively scale-normalize distance metrics, to robustly handle data outliers, to elucidate informative data points via regression models, to circumvent summary statistics via optimal transport based distances, and to avoid local optima in acceptance threshold sequences by predicting acceptance rate curves.
Further, we provide, besides previously existing support of Python and R, interfaces in particular to the Julia language, the COPASI simulator, and the PEtab standard.
\end{abstract}

\section{Introduction}

\begin{figure}[t!]
    \centering
    \includegraphics[width=0.99\textwidth]{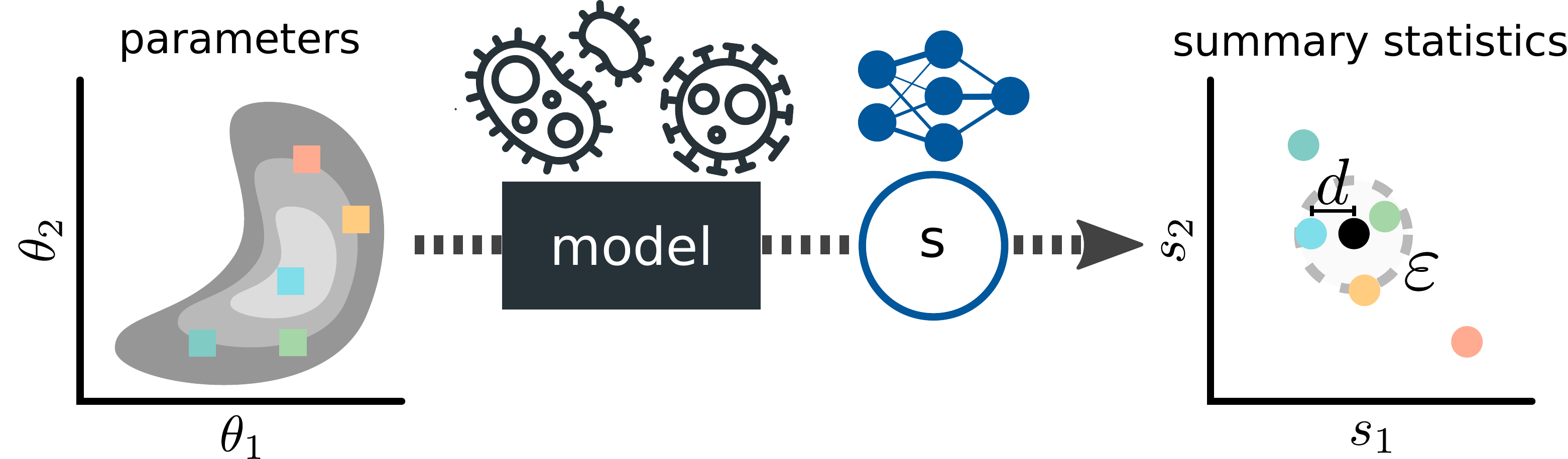}
    \caption{
    \textbf{Basic ABC algorithm.} Parameters $\theta\sim\pi(\theta)$ are sampled from the prior or a proposal distribution, and passed to a black-box model generating potentially stochastic simulated data according to the likelihood $y\sim\pi(y|\theta)$. These are optionally passed through a summary statistics function giving a low-dimensional representation $s(y)$. Summary statistics of simulated and observed data are compared via a distance metric $d$, and the underlying parameters accepted if the distance is below an acceptance threshold $\varepsilon$.
    }
    \label{fig:concept}
\end{figure}

Mathematical models are important tools to describe and study real-world systems, allowing to understand underlying mechanisms \citep{Gershenfeld1999, Kitano2002}.
They are commonly subject to unknown parameters that need to be estimated using observed data \citep{Tarantola2005}.
The Bayesian framework allows doing so by updating prior beliefs about parameters by the likelihood of data given parameters.
However, especially for complex stochastic models, evaluating the likelihood is often infeasible \citep{TavareBal1997, Wilkinson2009, JagiellaRic2017}.
Thus, likelihood-free methods such as ABC have been developed \citep{AndrieuRob2009, GourierouxMon1993, PriceDro2018, PritchardSei1999, BeaumontZha2002}.
ABC is widely applicable, as it only requires an executable ``black-box'' forward process model, simulating data given model parameters.
In a nutshell, ABC circumvents likelihood evaluation by accepting if a distance between simulated and observed data is below a threshold \citep{SissonFan2018Handbook} (Figure \ref{fig:concept}).
ABC is often combined with a sequential Monte-Carlo (ABC-SMC) scheme using importance sampling, which gradually reduces the acceptance threshold and thus improves the posterior approximation, while maintaining high acceptance rates \citep{SissonFan2007, DelMoralDou2006}.

While conceptually simple and widely applicable, ABC is computationally expensive, and its practical performance relies on a number of factors.
There exist several software packages implementing different algorithms in different languages, each with their own strengths, including notably in Python ABCpy \citep{DuttaSch2017} and ELFI \citep{KangasraasioLin2016}, in Julia GpABC \citep{TankhilevichIsh2020}, and in R EasyABC \citep{JabotFau2013}.
pyABC implements at its core an ABC-SMC scheme based on \citet{ToniStu2010} and facilicates robust and efficient inference for a broad spectrum of applications via robust methods and self-tuned choices of hyperparameters, reducing the need for manual tuning.
An article on core features of pyABC was previously published \citep{KlingerRic2018}, discussing in particular adaptive transition kernels \citep{FilippiBar2013}, population sizes \citep{KlingerHas2017}, and wall-time efficient parallelization via dynamic scheduling on multi-core and distributed systems.
pyABC is in use in a variety of fields, e.g.\ to model virus transmission on cellular \citep{ImleKum2019} and population level \citep{KerrStu2021}, neuron circuits \citep{BittnerPal2021}, cancer \citep{ColomHer2021}, gene expression \citep{CoulierHel2021}, axolotl regeneration \citep{CostaOts2021}, universe expansion \citep{BernardoSai2021}, cardiac electrophysiology \citep{CantwellYum2019}, and bee colonies \citep{MinucciCur2021}.

\section{Methods}

The methods discussed in the following have been newly implemented in pyABC (version 0.12), with details in the API documentation and Jupyter example notebooks accessible via the online documentation:

\textit{Variations in different data scales and robustness to outliers.}
In ABC, a distance metric is used to quantify differences between simulated and observed data.
When simulations for different data points vary on different scales, highly variable ones dominate the acceptance decision.
\citet{Prangle2017} introduces a method to, in an ABC-SMC framework, iteratively update distance weights to normalize contributions.
A further problem are outliers, i.e.\ errors in the measurement process that the model does not account for \citep{GhoshVog2012, MotulskyChr2003}.
\citet{SchaelteAla2021} show that an approach adapted from \citet{Prangle2017} can robustly identify outliers and reduce their impact.
The approaches by \citet{Prangle2017} and \citet{SchaelteAla2021} are now implemented in pyABC.

\textit{Identify informative data.} Instead of operating on the full data, in ABC often summary statistics, i.e.\ low-dimensional data representations, are employed \citep{BlumNun2013}.
A particular line of approaches uses as statistics the outputs of inverse regression models of parameters on data, e.g.\ via linear regression \citep{FearnheadPra2012}, neural networks \citep{JiangWu2017}, or Gaussian processes \citep{BorowskaGiu2021}.
In \citet{SchaelteHas2022Pre}, such approaches are combined with adaptive scale-normalization, and extended to achieve a higher-order posterior approximation. Further, inverse regression models are used to, instead of constructing summary statistics, inform robust sensitivity weights accounting for informativeness.
All of the above approaches are now implemented in pyABC, with regression models interfaced via scikit-learn \citep{scikitlearn2011}.

\textit{Accurate handling of noise.} The approximation error of ABC methods is often unclear.
\citet{Wilkinson2013} shows that ABC can be considered as giving exact inference under an implicit distance-induced noise model.
In \citet{SchaelteHas2020}, this insight is used to develop an efficient ABC-SMC based exact inference scheme in the presence of measurement noise. The framework is now integrated into pyABC.

\textit{Optimal transport distances.} Besides the above-mentioned adaptive distances, pyABC now in particular implements Wasserstein distances, which consider an optimal transport problem between distributions and may allow to circumvent the use of summary statistics \citep{BerntonJac2019}.

\textit{Acceptance threshold selection.} Efficiency and convergence of ABC-SMC algorithms further depend on the acceptance threshold sequence. \citet{SilkFil2013} discuss that common schemes based on quantiles of previous values \citep{DrovandiPet2011} can fail in the presence of local minima, and propose a method based on analyzing predicted acceptance rate curves.
pyABC now implements a modified version, using importance sampling instead of unscented transform to predict the acceptance rate as a function of the threshold.

\textit{Interoperability.} Not only algorithms, but also accessibility and interoperability determine the usefulness of a tool.
Besides natural support of Python, and previously established support of R, pyABC now also provides an efficient interface to models written in Julia \citep{BezansonEde2017}, to biochemical pathway models defined in SBML or COPASI format by using the COPASI toolbox \citep{HoopsSah2006}, and supports the PEtab inference standard \citep{SchmiesterSch2021}, currently only for the ODE simulator AMICI \citep{FroehlichWei2021}.
Finally, it allows to connect to models written in arbitrary languages and frameworks via file exchange.

\section{Availability and Documentation}

pyABC is being developed open-source under a 3-clause BSD license. The code, designed to be highly modular and extensible, is hosted on GitHub (\url{https://github.com/icb-dcm/pyabc}) and continuously tested.
Extensive documentation is hosted on Read the Docs (\url{https://pyabc.readthedocs.io}), including API documentation and numerous Jupyter notebooks containing tutorials, outlining features, and showcasing applications.

\section*{Acknowledgments}

We thank many collaboration partners and pyABC users for valuable input, in particular Frank Bergmann for the COPASI wrapper, and Elba Raimúndez for fruitful discussions.
This work was supported by the German Federal Ministry of Education and Research (BMBF)
(FitMultiCell/031L0159 and EMUNE/031L0293) and the German Research Foundation (DFG)
under Germany’s Excellence Strategy (EXC 2047 390873048 and EXC 2151 390685813).

\bibliography{main}

\end{document}